\documentclass[12pt]{article}

\usepackage{amsthm,graphicx,cite,braket}
\usepackage{epsf,latexsym}
\usepackage{amssymb,amsmath}
\usepackage{orcidlink}

\newcommand{\rmi}{{\rm i}}

\newcommand{\gr}{\Gamma _{\text{\tiny R}}}
\newcommand{\er}{E_{\text{\tiny R}}}
\newcommand{\tr}{\tau_{\text{\tiny R}}}

\newcommand{\mr}{m_{\text{\tiny R}}}
\newcommand{\pr}{p_{\text{\tiny R}}}
\newcommand{\pvec}{\vec{p}\,}

\begin{document}

\title{The Gamow and the Fermi Golden Rules}

\author{Rafael de la Madrid\,\orcidlink{0000-0001-9350-5371} \\
\small{\it Department of Physics, Lamar University,
Beaumont, TX 77710} \\
\small{E-mail: \texttt{rafael.delamadrid@lamar.edu}}}

\date{\small{May 23, 2024}}






\maketitle

\begin{abstract}
\noindent 
By using the fact that the Gamow states in the momentum representation
are square integrable, we obtain the differential and the total decay width
of a two-body, non-relativistic decay. The resulting Gamow Golden Rule is
well suited to describe both energy and angular decay distributions, and it
becomes the Fermi Golden Rule when the resonance is long-lived and far from
the energy threshold. We also show that the correct density of states and
phase space factors arise naturally from the Gamow Golden Rule. The upshot
is that the Gamow states and the Golden Rule can be combined into a unified
description of quantum resonances.

\end{abstract}


\newpage

\section{Introduction}
\label{sec:intro}

The decay energy spectra (invariant mass distributions) of resonances are
routinely measured in nuclear and particle physics. Typically, the number of
decays per unit of energy (mass) $dN/dE$ ($dN/dm$) is measured, see for example 
Refs.~\cite{MARTINSHAW,HIGGSA,HIGGSC,LOPEZSAAVEDRA,TETRA,
BABAR22,MONTEAGUDO,CMS24,LHCb24,BESIII24}. Angular decay
distributions such as $dN/d\hskip0.2mm {\cos \theta}$, $dN/dEd\Omega$ or 
$dN/d\phi$ are also common, see for example 
Refs.~\cite{Wehle:2016yoi,Aaboud:2018krd,LHCb-PAPER-2020-002,
cms20an,LHCb2021,WANGETAL,GAO24}.

Theoretically, decay energy spectra are usually described by way of
Fermi's Golden Rule, which provides the differential
decay width $\frac{d\overline{\Gamma}_{{\rm i}\to {\rm f}}}{dE_{\rm f}}$ 
for the decay from an 
energy eigenstate $|E_{\rm i}\rangle$ into a continuum 
of energy eigenstates $|E_{\rm f}\rangle$ whose density of states is 
$\rho (E_f)$~\cite{COHEN,NOTE1},
\begin{equation}
     \frac{d\overline{\Gamma}_{{\rm i}\to {\rm f}}}{dE_f}=
            2 \pi \rho (E_f)
          \left| \langle E_{\rm f}|V|E_{\rm i} \rangle \right |^{2} 
            \delta (E_{\rm i}-E_{\rm f}) \, ,
            \label{usualgr2}
\end{equation}
where $\langle E_{\rm f}|V|E_{\rm i} \rangle$ is the matrix element
of the interaction. The total decay width from the initial state to any state 
of the continuum is given by
\begin{equation}
   \overline{\Gamma}_{{\rm i}\to {\rm f}}= \int dE_{\rm f} \, 
      \frac{d\overline{\Gamma}_{{\rm i}\to {\rm f}}}{dE_f}=
          2 \pi \rho (E_{\rm f}=E_{\rm i})
          \left| \langle E_{\rm f}=E_{\rm i}|V|E_{\rm i} \rangle \right |^{2} 
             \, .  
            \label{usualgr3}
\end{equation}
Relativistically, the differential decay width for the decay of a resonance R 
into $n$ particles, 
${\rm R} \to 1+2 + \cdots + n$, is given by~\cite{PESKIN,GRIFFITHS}
\begin{equation}
     d\overline{\Gamma}  =   \frac{S}{2\mr} 
    |{\cal M}|^2 (2\pi)^4 \delta ^4 ( \pr -\sum_{i=1}^np_i)
     \prod_{j=1}^n \frac{1}{2E_j} 
      \frac{d^3\vec{p}_j}{(2\pi)^3} \, , 
        \label{relafgr}
\end{equation}
where $S$ is a statistical factor, ${\cal M}$ is the invariant matrix element 
of the interaction, and where the $m$'s, the 
$p$'s, the $\vec{p}\,$'s and the $E$'s are, respectively, the masses,
the four-momenta, the three-momenta and the energies of the particles.

Even though Eq.~(\ref{relafgr}) is supposed to be the relativistic
generalization of Eq.~(\ref{usualgr2}), those two equations have some notable
differences. Equation~(\ref{usualgr2}) utilizes an energy
basis and is the result of first-order perturbation theory, whereas 
Eq.~(\ref{relafgr}) uses a momentum basis and is derived from 
Quantum Field Theory under the only approximation that the resonance is 
long-lived~\cite{PESKIN}.

In non-relativistic quantum mechanics, the wave function contains all
the information necessary to calculate the probabilities for any possible 
outcome of any experiment. Because the Gamow (resonant) states have 
all the phenomenological properties to describe a resonance, it should be 
possible to use the Gamow states to account for Fermi's Golden Rule. In fact,
in Ref.~\cite{NPA15} we used the Gamow states and an energy basis to
obtain the differential and the total decay widths of a resonance. We also
showed that in the approximation that the resonance is sharp (long-lived), the
resulting Gamow Golden Rule bears a strong resemblance with Fermi's
Golden Rule, the main difference being that the density of states of
the Gamow Golden Rule is unity. The results of Ref.~\cite{NPA15} were
used to calculate the decay energy spectra of the resonances of
the delta-shell potential~\cite{NPA17}, and of a simplified model of
oxygen-25~\cite{NPA18}.

The formalism of Refs.~\cite{NPA15,NPA17,NPA18} has two main limitations. First,
it describes only energy (but not angular) decay distributions. Second, its
density of states is unity, whereas the density of states of Fermi's Golden
Rule is not unity. Therefore, it is not clear if the Gamow Golden Rule is truly
the same as Fermi's Golden Rule.

The purpose of this paper is to derive the Gamow Golden Rule using a
momentum basis, and to show that when the resonance is long-lived, the
Gamow Golden Rule becomes Fermi's Golden Rule. The momentum-basis Gamow
Golden Rule is the non-relativistic counterpart of Eq.~(\ref{relafgr})
and will allow us to describe both energy and angular decay
distributions. We
will also show that the Gamow Golden Rule has the correct density of states 
built into it. In particular, we will show that the expression obtained
in Ref.~\cite{NPA15} is indeed the Golden Rule when we use a
$\delta$-normalized energy basis for the decay products. 

The exponential blowup of the Gamow states in the position representation
seems to preclude the possibility to use them to obtain well-defined decay
distributions. However, we will see that, because the Gamow states in the
momentum representation are square integrable, we can use them to
obtain well-defined, normalizable decay distributions.

We will denote three similar quantities (resonant width, decay width,
and decay constant) by similar symbols ($\gr$, $\overline{\Gamma}$, and 
$\Gamma$), and it is helpful to describe them before we proceed with 
the derivation of the Gamow Golden Rule. First, $\gr$ 
is twice the imaginary part of the pole of the $S$-matrix,
$E_{\rm res}=\er -i\gr /2$. Physically, $\gr$ is the inverse of the lifetime
$\tr =\hbar /\gr$, and it is also a measure of the spread of the resonant
energies ($\er$ is just the mean value of the possible energies a 
resonance can have). Second, $\overline{\Gamma}$ is the total decay width,
has units of energy, and is a measure of how strongly the resonance couples
to the continuum. The decay width $\overline{\Gamma}$ divided by $\hbar$ can 
also be thought of as a decay rate~\cite{NPA15}. Third, $\Gamma$ is 
a dimensionless quantity that can also be used to characterize the 
strength of the coupling between the resonance and the continuum. As shown 
in Ref.~\cite{NPA15}, these three quantities are related by
\begin{equation}
       \overline{\Gamma} = \gr \Gamma \, .
           \label{dewvsdcta}
\end{equation}
Both $\overline{\Gamma}$ and $\Gamma$ have differential counterparts
that are related by
\begin{equation}
        \frac{d\overline{\Gamma}(E)}{dE} = \gr \frac{d\Gamma(E)}{dE} \, .
           \label{dewvsdctb} 
\end{equation}
The resonant width $\gr$ does not have a differential version, because if
it did, we could define a differential version of the lifetime, which
would imply that the lifetime of a resonance is energy-
and/or angular-dependent, something
for which there is no experimental evidence. 

The structure of the paper is as follows. In Sec.~\ref{sec:gsdervmomentbasis}, 
we derive the Gamow Golden Rule in the momentum basis, and argue
that it can describe both angular and energy decay distributions. In
Sec.~\ref{sec:gsdervangmomentbasis}, we derive the Gamow Golden
Rule in the angular momentum basis. In Sec.~\ref{sec:relationship}, we 
show that the momentum and the angular momentum bases yield the same decay
constant when the potential is spherically symmetric. In
Sec.~\ref{sec:originalfgr}, we show that when the resonance is 
long-lived, the (exact) Gamow Golden Rule becomes the
(approximate) Fermi Golden Rule. In Sec.~\ref{sec:dos}, we show that 
the Gamow Golden Rule has the correct density of final states. In 
Sec.~\ref{sec:phasespace}, we discuss the 
differences between the phase space factors of the Gamow and of
the Fermi Golden Rules. In Sec.~\ref{sec:siogs}, we explain how the 
Lippmann-Schwinger equation of a Gamow state can be used to obtain
well-defined, square-integrable Gamow eigenfunctions in the momentum 
representation, even though the Gamow eigenfunctions blow up exponentially
in the position representation. Section~\ref{sec:conclusions} contains
our conclusions.

\section{The Gamow Golden Rule in the momentum basis}
\setcounter{equation}{0}
\label{sec:gsdervmomentbasis}

Let us consider a spinless particle of mass $m$ impinging on a potential
$V$ (or the relative motion of two particles of reduced mass $m$ interacting
through the potential $V$). Let $H=H_0+V$ be the Hamiltonian, where $H_0$ is the
free Hamiltonian. The momentum eigenstates of the free Hamiltonian will 
be denoted by $|\pvec\rangle$,
\begin{equation}
       H_0|\vec{p}\, \rangle = E_p |\vec{p} \, \rangle = \frac{p^2}{2m} 
                |\vec{p}\, \rangle \, .
         \label{freehe}
\end{equation}
The stationary states of scattering theory satisfy the Lippmann-Schwinger 
equation~\cite{TAYLOR},
\begin{equation}
       |\vec{p}\,^{+}\rangle = |\vec{p}\, \rangle + 
        \frac{1}{E-H_0+ i 0}V|\vec{p}\,^{+}\rangle \, .
         \label{lipsch}
\end{equation}
This equation is equivalent to the time-independent Schr\"odinger equation,
\begin{equation}
       H|\vec{p}\,^{+}\rangle = E_p |\vec{p}\, ^{+}\rangle = \frac{p^2}{2m} 
                |\vec{p}\, ^{+}\rangle \, ,
         \label{fulshdhe}
\end{equation}
subject to the boundary condition that, in the position representation, 
the eigenfunction $\langle \vec{x}|\vec{p}\,^{+}\rangle$ represents an 
incoming plane wave in the distant past and an outgoing spherical wave
in the distant future,
\begin{equation}
  \langle \vec{x}|\vec{p}\,^{+} \rangle \xrightarrow[|\vec{x}|\to \infty]{}
       (2\pi \hbar)^{-3/2}\left[ e^{i\vec{p}\cdot\vec{x}/\hbar}+
            f(\vec{p} \to p\hat{x} )
                        \frac{e^{ipr/\hbar}}{r}\right]  \, , 
          \label{asymptotic+}
\end{equation}
where $f(\vec{p} \to p\hat{x} )$ is the scattering amplitude.

We will denote the resonant (Gamow) state by $|p_{\rm res} \rangle$, where 
$\frac{p_{\rm res}^2}{2m} = E_{\rm res}$~\cite{NOTE2}. Instead 
of the Lippmann-Schwinger equation~(\ref{lipsch}),
the resonant states satisfy~\cite{MONDRA,WOLF}
\begin{equation}
    |p_{\rm res}\rangle = 
           \frac{1}{E_{\rm res} -H_0 +\rmi 0}V |p_{\rm res}\rangle \, . 
         \label{inteGam0}
\end{equation}
This integral equation is equivalent to the time-independent 
Schr\"odinger equation
\begin{equation}
       H|p_{\rm res}\rangle = E_{\rm res} |p_{\rm res} \rangle
       = \frac{p_{\rm res}^2}{2m} |p_{\rm res} \rangle \, ,
       \label{timeindschgs}
\end{equation}
subject to a purely outgoing boundary condition,
\begin{equation}
  \langle \vec{x}|p_{\rm res} \rangle \xrightarrow[|\vec{x}|\to \infty]{}
                      g(\hat{x},p_{\rm res})  \frac{e^{ip_{\rm res}r/\hbar}}{r} \, ,
       \label{pobc}
\end{equation}
where the function $g(\hat{x},p_{\rm res})$ does not depend on $r$. It 
is easy to show that Eq.~(\ref{timeindschgs}) follows from
Eq.~(\ref{inteGam0}). It is easy to understand why Eq.~(\ref{pobc})
follows from Eq.~(\ref{inteGam0}). Formally, Eq.~(\ref{inteGam0}) is
the same as the Lippmann-Schwinger equation~(\ref{lipsch}) but without
the non-homogeneous term $|\pvec\rangle$. Hence, the
asymptotic behavior of $\langle \vec{x}|p_{\rm res}\rangle$ 
should be the same as that of $\langle \vec{x}|\vec{p}^{\, +}\rangle$ 
but without the contribution of the plane wave 
$(2\pi \hbar)^{-3/2} e^{i\vec{p}\cdot\vec{x}/\hbar}$. Removing the contribution 
of the plane 
wave from the asymptotic behavior~(\ref{asymptotic+}) leaves the
outgoing wave $e^{ip_{\rm res}r/\hbar}/r$ multiplied by a factor that is 
independent of $r$.

Although the Gamow eigenfunctions blow up at infinity due to their 
asymptotic behavior, it is possible to normalize them to unity by 
using a regularization such as Zeldovich's regulator, complex scaling, or 
analytic continuation~\cite{GSM,MAREK,ROMO}. With such normalization, the Gamow 
eigenfunctions $\langle \vec{x}|p_{\rm res}\rangle$ have dimensions of 
(length)$^{-3/2}$, just like normalized bound states.

Let us now derive the expression for the decay spectrum. If we take
the inner product of Eq.~(\ref{inteGam0}) with $\langle \pvec|$
and use Eq.~(\ref{freehe}), we obtain
\begin{equation}
   \langle \pvec |p_{\rm res}\rangle = 
           \langle \pvec|\frac{1}{E_{\rm res} -H_0 +\rmi 0}V |p_{\rm res}\rangle 
= \frac{1}{E_{\rm res} -E_p} \langle \pvec|V |p_{\rm res}\rangle\, . 
         \label{inteGamp0}
\end{equation}
If $|\pvec\rangle$ is delta 
normalized and $|p_{\rm res}\rangle$ is normalized to unity, then
$\langle \pvec |p_{\rm res}\rangle$ has units
of momentum$^{-3/2}$. In fact, as we will see in Sec.~\ref{sec:siogs}, 
$\langle \pvec |p_{\rm res}\rangle$ can be thought of as the momentum 
representation  of the Gamow state. Therefore,
$|\langle \pvec |p_{\rm res}\rangle|^2$
can be interpreted as the momentum probability density for 
the decay of a resonance
of complex momentum $p_{\rm res}$ to a stable particle of 
momentum $\pvec$,
\begin{equation}
    \frac{dP(p_{\rm res} \to \pvec)}{d^3p} =
      |\langle \pvec |p_{\rm res}\rangle|^2 \, .
    \label{momenproden}
\end{equation}
It is natural to identify this probability density with the decay 
spectrum $d\Gamma/d^3p$ of a resonance decaying into a stable particle of 
momentum $\pvec$,
\begin{equation}
      \frac{d\Gamma (p_{\rm res} \to \pvec)}{d^3p} \equiv 
\frac{dP(p_{\rm res} \to \pvec)}{d^3p}    \, .
    \label{momenprodenbis}
\end{equation}
Combining Eqs.~(\ref{inteGamp0}), (\ref{momenproden}) 
and~(\ref{momenprodenbis}) yields
\begin{equation}
   \frac{d\Gamma (p_{\rm res} \to \pvec )}{d^3p} =
\frac{1}{(E_p-\er)^2+(\gr/2)^2}  
         |\langle \pvec |V|p_{\rm res}\rangle|^2 \, .
    \label{momenproden2}
\end{equation}
Hence, the probability 
per unit of momentum volume for a resonance to decay into a stable
particle of momentum $\vec{p}$ is given by the Lorentzian lineshape
times the absolute value squared of the matrix element of the interaction
between the Gamow state and the free scattering state.

Because $d^3p=p^2dp\,d\Omega_{\hat{p}}= m\sqrt{2mE_p}\, dE_p\,d\Omega_{\hat{p}}$, 
we can write Eq.~(\ref{momenproden2}) as
\begin{equation}
  \frac{d\Gamma(p_{\rm res} \to \pvec)}{dE_p\,d\Omega_{\hat{p}}} =
 m\sqrt{2mE_p}\frac{1}{(E_p-\er)^2+(\gr/2)^2}  
         |\langle \pvec |V|p_{\rm res}\rangle|^2 \, .
    \label{momenproden2a}
\end{equation}
This quantity is appropriate to describe experiments where counts (events) 
per unit of energy and solid angle, $dN/dEd\Omega$, are measured. 
By integrating 
Eq.~(\ref{momenproden2a})
over $d\Omega_{\hat p}$, we obtain the decay energy spectrum,
\begin{equation}
     \frac{d\Gamma (p_{\rm res} \to E_p)}{dE_p} = 
 m\sqrt{2mE_p}\frac{1}{(E_p-\er)^2+(\gr/2)^2}  
        \int d\Omega_{\hat p} \, 
           |\langle \pvec |V|p_{\rm res}\rangle|^2 \, .
    \label{momenproden2b}
\end{equation}
This quantity is appropriate to describe experiments where counts (events) 
per unit of energy, $dN/dE$, are measured. By writing 
$d\Omega_{\hat{p}}=d(\cos \theta) d\phi$ in Eq.~(\ref{momenproden2a}),
and by carrying out the appropriate integrations, we can describe
angular decay distributions such as $dN/d\hskip0.2mm {\cos \theta}$ 
and $dN/d\phi$.

The total decay constant for decay of the resonance into the continuum 
is given by
\begin{equation}
    \Gamma = \int dE_p\, d\Omega_{\hat p} \,
 m\sqrt{2mE_p}\frac{1}{(E_p-\er)^2+(\gr/2)^2}  
           |\langle \pvec |V|p_{\rm res}\rangle|^2 \, .
    \label{momenproden2bto}
\end{equation}

In the way we wrote them, neither the theoretical nor the experimental 
differential decay distributions are normalized. Once they are normalized, 
the correct relationship between them is
\begin{equation}
     \frac{1}{\Gamma} \frac{d\Gamma}{dE_p} =
     \frac{1}{\overline{\Gamma}} \frac{d\overline{\Gamma}}{dE_p}
         =  \frac{1}{N} \frac{dN}{dE_p} \, ,
     \label{normpdis}
\end{equation}
where we have used Eqs.~(\ref{dewvsdcta}) and~(\ref{dewvsdctb}). Similar
expressions hold for the normalized angular decay 
distributions. Equation~(\ref{normpdis}) shows, in particular, that we 
can use either the decay width $\overline{\Gamma}$ or the decay constant 
$\Gamma$ to describe normalized decay distributions.

As shown in Ref.~\cite{NPA17}, the Gamow Golden Rule can also describe
overlapping resonances. If instead of a single resonance
$|p_{\rm res}\rangle$, we have for example two resonances
$|p_{\rm res,1}\rangle$ and $|p_{\rm res,2}\rangle$ that contribute to the decay
distribution, then the system is described by a superposition
of two Gamow states, $c_1 |p_{\rm res,1}\rangle + c_2 |p_{\rm res,2}\rangle$. 
Replacing the single Gamow state $|p_{\rm res}\rangle$ by such
superposition in Eqs.~(\ref{inteGamp0})-(\ref{momenprodenbis})
yields~\cite{NPA17}
\begin{eqnarray}
       \frac{d \Gamma }{d^3p} = |c_1|^2 
       \frac{1}{(E_p-E_1)^2+(\Gamma _1/2)^2}|\langle \pvec|V|p_{\rm res,1}
                \rangle |^2+
   |c_2|^2 \frac{1}{(E_p-E_2)^2+(\Gamma _2/2)^2}|\langle \pvec|V|p_{\rm res,2} \rangle |^2
        \nonumber \\
        + 2 \, \text{Re} \left( c_1c_2^* \frac{1}{(E_p -E_1)+i\Gamma_1/2}
         \frac{1}{(E_p-E_{2})-i\Gamma_2/2}
         \langle \pvec |V|p_{\rm res,1} \rangle
         \langle \pvec |V|p_{\rm res,2} \rangle ^* \right) \, .
         \hskip0.8cm
      \label{ddconsvdd3}
\end{eqnarray}
Thus, when two resonances interfere, the Gamow Golden Rule yields a
decay spectrum that consists of the sum of the individual decay spectra
plus an interference term. As pointed out in Ref.~\cite{NPA17}, the
Fermi Golden Rule cannot account for such interference term, since
the interference of two delta functions should be zero, and therefore
the Gamow Golden Rule is more appropriate to describe resonance interference
than the Fermi Golden Rule.

The Gamow Golden Rule applies equally well to sharp (long-lived) and to
broad (short-lived) resonances, as long as there is a pole of the
$S$-matrix (and therefore a Gamow state) associated with the resonance. Thus,
qualitatively, the Gamow Golden Rule does not distinguish between
short-lived and long-lived resonances. Quantitatively however, a long-lived
resonance has a sharp decay distribution that can be approximately described by
Fermi's Golden Rule, whereas a short-lived resonance has a broader decay
distribution that is not well approximated by Fermi's Golden Rule.

When a single-channel resonance is close to the energy threshold, the
Lorentzian gets distorted on the side close to the threshold, which produces
an overall distortion of the decay distribution~\cite{NPA17}. In the
extreme case that the resonance becomes a virtual state, the decay
distribution looks like half a peak (with the other half ``missing'' below the
threshold)~\cite{NPA17}. Although the Gamow Golden Rule for multi-channel
resonances has not been derived yet, it is likely that when a
multi-channel resonance is close to an inelastic threshold, the shape of
the decay distribution gets distorted by both the presence of the threshold
and the interference with other channels.

\section{The Gamow Golden Rule in 
the angular momentum basis}
\setcounter{equation}{0}
\label{sec:gsdervangmomentbasis}

If the interaction is spherically symmetric, it is convenient to use an angular
momentum basis. In such basis, the resonant states can be written as
\begin{equation}
         \langle \vec{x}|E_{\rm res},l,m \rangle =
           \frac{u_l(r;E_{\rm res})}{r}Y_l^m(\hat{x} ) \, ,
\end{equation}
where $Y_l^m(\hat{x})$ are the spherical harmonics, and
where the (regularized) normalization of $u_l(r;E_{\rm res})$ is such that
\begin{equation}
       \int_0^{\infty}dr \, [u_l(r;E_{\rm res})]^2 = 1 \, .
            \label{normalization}
\end{equation}
The delta-normalized eigenfunctions of the free Hamiltonian
in the angular momentum basis are~\cite{TAYLOR}
\begin{equation}
       \langle \vec{x}|E,l,m\rangle = 
             \frac{\psi_l^0(r;E)}{r} Y_l^m(\hat{x}) =
              i^l \sqrt{\frac{2m}{\pi \hbar p}}
                \frac{1}{r} \hat{j}_l(pr/\hbar) Y_l^m(\hat{x})  \, ,
      \label{andmbinpos}
\end{equation}
where $\hat{j}_l(x)$ are the Ricatti-Bessel functions.

Using the angular momentum basis, the Lippmann-Schwinger equation of
a resonant state can be written as
\begin{equation}
         |E_{\rm res}, l,m \rangle = 
\frac{1}{E_{\rm res} -H_0 +\rmi 0}V|E_{\rm res}, l , m\rangle 
                 \, .
          \label{inteGam0ener}
\end{equation}
By taking the inner product of Eq.~(\ref{inteGam0ener}) with the eigenbra 
$\langle E,l',m'|$, and by using the fact that $\langle E,l',m'|$
is an eigenvector of $H_0$ with eigenvalue $E$, we obtain
\begin{equation}
   \langle E,l',m' |E_{\rm res},l,m\rangle = \frac{1}{E_{\rm res} -E} 
        \langle E,l',m' |V |
          E_{\rm res},l,m \rangle \, . 
         \label{inteGam2}
\end{equation}
Since $V$ is spherically symmetric, 
\begin{eqnarray}
        \langle E,l',m' |V | E_{\rm res},l,m \rangle
         &=& \int_0^{\infty} r^2 dr \, \frac{\psi_{l'}^0(r,E)^*}{r} V(r)
           \frac{u_l(r;E_{\rm res})}{r} 
            \int d\Omega_{\hat{x}} Y_{l'}^{m'}(\hat{x})^* Y_l^m(\hat{x})
             \nonumber \\
        & =&
          \int_0^{\infty} dr\, \psi_{l'}^0(r,E)^* V(r) u_l(r;E_{\rm res}) \, 
          \delta_{l,l'}\delta_{m,m'} \nonumber \\
        & \equiv & \langle E|V|E_{\rm res}\rangle _l \
          \delta_{l,l'}\delta_{m,m'}  \, ,  
           \label{inteGam2a}
\end{eqnarray}
where the reduced matrix element $\langle E|V|E_{\rm res}\rangle _l$ does not 
depend on the orbital
magnetic quantum number $m$. Combining Eqs.~(\ref{inteGam2}) 
and~(\ref{inteGam2a}), and identifying the $l$th partial-wave differential 
decay constant with the absolute value squared 
of $\langle E, l',m'|E_{\rm res}, l, m \rangle$, we obtain
\begin{equation}
\frac{d\Gamma_l (E_{\rm res}, l,m \to E,l',m')}{dE} =  
           \frac{1}{(E-\er)^2+(\gr/2)^2} |\langle E|V |E_{\rm res} \rangle _l|^2 
            \delta_{l,l'}\delta_{m,m'}   \, .             
         \label{inteGam3}
\end{equation}
Hence, the probability density for decay of an $l$th-wave resonance into the 
continuum with energy $E$ is given by the Lorentzian lineshape times 
the absolute value squared of the (reduced) matrix element of the
interaction. The $l$th partial-wave decay constant is
\begin{equation}
       \Gamma_l  = \sum_{l',m'}\int dE \, 
             \frac{d\Gamma}{dE}(E_{\rm res}, l,m \to E,l',m')
             = \int dE
       \frac{1}{(E-\er)^2+(\gr/2)^2} |\langle E|V |E_{\rm res} \rangle _l|^2 
           \label{inteGam4}
\end{equation}  
We can interpret $\Gamma_l$ as a dimensionless quantity that estimates
the strength of the coupling between the resonance and the 
$l$-th partial-wave continuum.

\section{The relationship between the momentum and 
the angular momentum Golden Rules}
\setcounter{equation}{0}
\label{sec:relationship}

To double check the consistency of our results, and to be able to
clarify the role that the density of states plays in the Golden Rule 
(see Sec.~\ref{sec:dos}), we are going to
show that when the potential is spherically symmetric, 
the Gamow Golden Rule in the momentum basis yields
the same decay constant as in the angular momentum basis.

Let us first recall the expansion of plane waves in terms of spherical
harmonics~\cite{TAYLOR}:
\begin{eqnarray}
      \frac{1}{(2\pi\hbar)^{3/2}} e^{i\vec{p}\cdot \vec{x}/\hbar} &=&
    \sum_{l',m'} i^{l'}\sqrt{\frac{2}{\pi \hbar}} \frac{1}{pr} 
         \hat{j}_{l'}(pr/\hbar)
          Y_{l'}^{m'}(\hat{x}) Y_{l'}^{m'}(\hat{p})^*   \nonumber \\
        &=&
    \sum_{l',m'} \sqrt{\frac{1}{m p}} \frac{\psi_{l'}^0(r;E)}{r} 
          Y_{l'}^{m'}(\hat{x}) Y_{l'}^{m'}(\hat{p})^*  \, ,
\end{eqnarray}
where we have used Eq.~(\ref{andmbinpos}). The matrix element of the
interaction becomes
\begin{eqnarray}
     \langle \pvec | V | p_{\rm res}\rangle &=& 
          \int d^3x \,   \frac{1}{(2\pi\hbar)^{3/2}} e^{-i\vec{p}\cdot \vec{x}/\hbar}
           V(r) \frac{u_l(r;E_{\rm res})}{r}Y_l^m(\hat{x} ) \nonumber    \\
           &=& 
          \int r^2 dr d\Omega_{\hat{x}} \, 
    \sum_{l',m'} \sqrt{\frac{1}{m p}} \frac{\psi_{l'}^0(r;E)^*}{r} 
          Y_{l'}^{m'}(\hat{x})^* Y_{l'}^{m'}(\hat{p}) 
           V(r) \frac{u_l(r;E_{\rm res})}{r}Y_l^m(\hat{x} ) \nonumber    \\
   &=& \sqrt{\frac{1}{m p}}
    \sum_{l',m'}  \int dr  \,  \psi_{l'}^0(r;E)^* 
              V(r) u_l(r;E_{\rm res}) Y_{l'}^{m'}(\hat{p})
       \int d\Omega_{\hat{x}} Y_{l'}^{m'}(\hat{x})^* Y_l^m(\hat{x} ) \nonumber    \\
     &=&  \sqrt{\frac{1}{m p}}
    \langle E|V|E_{\rm res}\rangle _l \, Y_{l}^{m}(\hat{p}) \, .
    \label{maelpte}
\end{eqnarray}
Substitution of Eq.~(\ref{maelpte}) into Eq.~(\ref{momenproden2bto})
gives
\begin{eqnarray}
    \Gamma &=& \int dE_p\, d\Omega_{\hat p} \,
 m\sqrt{2mE_p}\frac{1}{(E_p-\er)^2+(\gr/2)^2}  
       \frac{1}{m p}
    |\langle E|V|E_{\rm res}\rangle _l|^2 \, |Y_{l}^{m}(\hat{p})|^2           
      \nonumber \\
   &=& \int dE_p \, \frac{1}{(E_p-\er)^2+(\gr/2)^2}  
    |\langle E|V|E_{\rm res}\rangle _l|^2   \, , 
    \label{momenproden2btobis}
\end{eqnarray}
which is the $l$th partial decay constant of Eq.~(\ref{inteGam4}), as
it should be.

\section{Long-lived resonances}
\setcounter{equation}{0}
\label{sec:originalfgr}

When the resonance is long-lived (sharp) and far from the energy threshold, we
can replace the Lorentzian by the delta function,
\begin{equation}
     \lim _{\frac{\gr}{\er} \to 0} \frac{\gr/(2\pi)}{(E-\er)^2+(\gr/2) ^2}
     = \delta(E-\er) \, .
          \label{lgode}
\end{equation}
Substitution of this equation into Eq.~(\ref{momenproden2a}) gives
\begin{equation}
  \frac{d\Gamma(p_{\rm res} \to \pvec)}{dE_p\,d\Omega_{\hat{p}}} = 
      \frac{2\pi}{\gr} m\sqrt{2mE_p}  
         |\langle \pvec |V|p_{\rm res}\rangle|^2 \delta (E_p -\er) \, .
         \label{momenproden2as4}
\end{equation}
We can now use Eq.~(\ref{dewvsdctb}) to obtain the differential decay
width for long-lived resonances,
\begin{equation}
 \frac{d\overline{\Gamma}(p_{\rm res} \to \pvec)}{dE_p\,d\Omega_{\hat{p}}} = 
      2\pi m\sqrt{2mE_p}  
         |\langle \pvec |V|p_{\rm res}\rangle|^2 \delta (E_p -\er) \, .
    \label{momenproden2as4bis}
\end{equation}
As we are going to see in the next section, Eq.~(\ref{momenproden2as4bis})
becomes the standard Fermi Golden Rule when the 
scattering plane waves $|\vec{p}\rangle$ are box normalized. Hence, the (exact)
Gamow Golden Rule becomes the (approximate) Fermi Golden Rule when
the resonance is long-lived.

Phenomenologically, the main difference between the Fermi Golden Rule,
Eqs.~(\ref{momenproden2as4}) and~(\ref{momenproden2as4bis}), and the
Gamow Golden Rule, Eq.~(\ref{momenproden2a}), is that the Gamow Golden Rule
reproduces the Lorentzian bumps of decay energy spectra, whereas the Fermi 
Golden Rule replaces the bumps by sharp spikes at the
mean value of the resonant energy. 

In the Fermi Golden Rule, the energy delta function is 
supposed to explicitly enforce energy conservation. Because in the
Gamow Golden Rule the energy delta function is replaced by the 
Lorentzian, it may appear that the Gamow Golden Rule violates
conservation of energy. The reason why the Gamow Golden Rule does not
include the energy delta function is that, experimentally, resonances are not
always produced at the mean resonant energy $\er$, the energies of
individual resonances have a spread (measured by $\gr$) around $\er$. For
example, Higgs bosons produced in proton-proton 
collisions~\cite{HIGGSA,HIGGSC} do not all have 
the same 125 GeV mass, their masses have a spread. However,
when an individual Higgs boson decays into, for example, two photons,
the sum of the energies of those two photons is exactly the same as the 
energy of the ``parent'' Higgs. Because the Golden Rule describes
the decay of a large number of resonances that are produced at various
energies, energy conservation is not explicitly enforced. Only when the 
resonance is long-lived and we can assume that all the resonances are
produced at the same mean energy $\er$, we can use a delta function to
enforce energy conservation explicitly.

\section{Density of states}
\setcounter{equation}{0}
\label{sec:dos}

Both the Gamow and the Fermi Golden Rules depend on how the wave 
functions that appear in the matrix element of the interaction
are normalized. We are going
to see in this section that the density of states of 
Fermi's Golden Rule arises naturally from the
Gamow Golden Rule when the plane
waves of the stable particles are box-normalized instead of delta-normalized.

The delta- and the box-normalized plane waves are given, 
respectively, by
\begin{eqnarray}
      && \langle \vec{x}|\pvec \rangle  = 
             \frac{1}{(2\pi \hbar)^{3/2}}
         e^{i\vec{p}\cdot \vec{x}/\hbar} \, , \\
      &&  \langle \vec{x}|\pvec {\rangle _{\rm box}} = \frac{1}{l^{3/2}}
         e^{i\vec{p}\cdot \vec{x}/\hbar} \, , 
\end{eqnarray}
where $l^3$ is the volume of the box where we confine our system. The 
matrix element of the interaction can then be written as
\begin{eqnarray}
      \langle \pvec | V | p_{\rm res}\rangle &=& 
      \int d^3\!x \,  \frac{1}{(2\pi \hbar)^{3/2}}
         e^{-i\vec{p}\cdot \vec{x}/\hbar} V(\vec{x}) \psi_{\rm res}(\vec{x},p_{\rm res}) 
         \nonumber \\
       &=&  \frac{l^{3/2}}{(2\pi \hbar)^{3/2}} 
         \int d^3\!x \,  \frac{1}{l^{3/2}}
         e^{-i\vec{p}\cdot \vec{x}/\hbar} V(\vec{x}) \psi_{\rm res}(\vec{x},p_{\rm res}) 
         \nonumber \\
       &\equiv &  \frac{l^{3/2}}{(2\pi \hbar)^{3/2}} \
           {_{\rm box} \langle} \pvec | V | p_{\rm res}\rangle
          \label{deltvbox}
\end{eqnarray}
By plugging Eq.~(\ref{deltvbox}) into Eq.~(\ref{momenproden2b}), we obtain
\begin{eqnarray}
  \frac{d\Gamma(p_{\rm res} \to E_p)}{dE_p} &=&
\frac{l^{3}}{(2\pi \hbar)^{3}}
 m\sqrt{2mE_p}d\Omega_{\hat{p}} \frac{1}{(E_p-\er)^2+(\gr/2)^2}  \
         \left| {_{\rm box}\langle} \pvec |V|p_{\rm res}\rangle \right|^2 
          \nonumber \\
         &=& \rho (E) \frac{1}{(E_p-\er)^2+(\gr/2)^2}  \
         \left| {_{\rm box}\langle} \pvec |V|p_{\rm res}\rangle \right|^2 \, ,
    \label{boxvdemomenproden}
\end{eqnarray}
where $\rho (E)$ is the density of states reported in the 
literature~\cite{GRIFFITHS2},
\begin{equation}
           \rho (E) = \frac{l^{3}}{(2\pi \hbar)^{3}}
 m\sqrt{2mE_p}d\Omega_{\hat{p}} \, .
    \label{dosgriffiths}
\end{equation}

When the resonance is long-lived, we can use Eqs.~(\ref{dewvsdctb}),
(\ref{lgode}) and (\ref{boxvdemomenproden}) to obtain the
differential decay width,
\begin{equation}
  \frac{d\overline{\Gamma}(p_{\rm res} \to E_p)}{dE_p} =
        2\pi \rho (E)
           \left| {_{\rm box}\langle} \pvec |V|p_{\rm res}\rangle \right|^2 \, ,
    \label{boxvdemomenprodenf}
\end{equation}
which is just Fermi's Golden Rule. Hence, the density of states of
Fermi's Golden Rule appears automatically in the Gamow Golden Rule by 
changing the normalization of the final states appropriately. Note, in
particular, that when we use an angular momentum (or an energy) basis such 
that its elements are delta-normalized, the resulting density of states is 
unity (see Eq.~(\ref{inteGam3}) and Ref.~\cite{NPA15}), even though we 
are still dealing with the same Golden Rule as when we have an explicit 
density of states.

\section{Phase space factors}
\setcounter{equation}{0}
\label{sec:phasespace}

When a resonance decays to $n$ stable particles, the general formula to 
calculate the non-relativistic density of states 
$\rho_n(E)$ in terms of phase space factors is~\cite{SKJEGGESTAD}
\begin{equation}
     \rho_n(E) = \int \prod_{j=1}^n  
      d^3\vec{p}_j \, \delta ( \er -\sum_{i=1}^nE_i) \, 
                  \delta ^{(3)} (\vec{p}_{\rm res} - \sum_{i=1}^n\vec{p}_i) \, ,
    \label{psfnr}
\end{equation}
where the delta functions enforce energy- and momentum-conservation 
explicitly. Those delta functions, however, do not appear in the
Gamow Golden Rule. In Sec,~\ref{sec:originalfgr}, we saw that the 
reason why
the energy-conservation delta function does not appear in the Gamow Golden
Rule is that the energies of resonances produced in collision experiments
have a spread (quantified by $\gr$) around the mean energy $\er$, and 
therefore we should not enforce that the energy of the decay products 
is always equal to
$\er$, unless the particle is long-lived and the energy of the resonances
varies little from $\er$. Since the energies of the resonances have a 
spread, the magnitude of their momenta also have a spread, and therefore
the Gamow Golden Rule does not include a delta function to
enforce momentum conservation. However, in each individual decay event,
both energy and momentum are conserved. 

Not only the magnitude of the momentum of the resonance changes from one 
decay event to another, its direction also changes. Hence, we cannot associate
with a resonance a momentum $\vec{p}_{\rm res}$ that has a specific direction  
(this is why we denote the Gamow state by $|p_{\rm res}\rangle$ 
instead of $|\vec{p}_{\rm res}\rangle$). There are two reasons for this. First,
$|p_{\rm res}\rangle$ is not an eigenvector of the momentum operator, and 
therefore cannot be associated with a specific direction in space. Measurement
of the momentum of the resonance for an individual decay event yields
a specific momentum, but a collection of such measurements would have
a statistical spread. (This is similar
to bound states, which are not eigenvectors of the momentum operator and 
therefore do not have a momentum pointing in a specific direction.)
Second, the Lippmann-Schwinger
equation of a Gamow state~(\ref{inteGam0}) describes a particle that
in the distant future does not have a momentum pointing in a specific
direction, because in the distant future the particle is described by
a spherical wave $g(\hat{x},p_{\rm res})\frac{e^{ip_{\rm res}r/\hbar}}{r}$, and
therefore the momentum of the resonance in each decay event can point in 
any radial direction with probability $|g(\hat{x},p_{\rm res})|^2$. This 
is similar to the Lippmann-Schwinger equation~(\ref{lipsch}), which although 
in the infinite past describes a particle with well-defined momentum 
$\vec{p}$, in the distant future describes a particle whose momentum
can point in any direction with probability
$|f(\vec{p}\to p\hat{x})|^2$, because long after the collision, the 
particle is described by an outgoing spherical wave
$f(\vec{p}\to p\hat{x}) \frac{e^{ipr/\hbar}}{r}$.

\section{Square integrability of the Gamow states in the 
momentum representation}
\setcounter{equation}{0}
\label{sec:siogs}

The main assumption behind the Gamow Golden Rule is that the
differential decay constant is given by the absolute value squared of the
Gamow state in the momentum representation, see 
Eqs.~(\ref{momenproden}) and~(\ref{momenprodenbis}). However, how
can we define the momentum representation of a wave function that in the
position representation blows up exponentially? It is a remarkable fact 
that the Lippmann-Schwinger equation of the Gamow states~(\ref{inteGam0}) 
yields Gamow eigenfunctions that are square integrable in the momentum 
representation, if the momentum representation is obtained by
analytic continuation, as shown in Ref.~\cite{MONDRA}. Essentially,
to find Gamow states by solving the integral equation~(\ref{inteGam0}), one
needs first to solve Eq.~(\ref{inteGam0}) in the upper half plane 
of the first Riemann sheet (first quadrant in the momentum plane), and then
continue the result to the resonant pole in the lower half plane of
the second Riemann sheet (fourth quadrant in the momentum 
plane)~\cite{MONDRA}. To specify this process, Eq.~(\ref{inteGam0}) 
can be written as~\cite{MONDRA}
\begin{equation}
    |p_{\rm res}\rangle = \left[
           \frac{1}{q^2/2m -H_0}V |q\rangle \right]_{q= p_{\rm res}} \, ,
         \label{inteGam08}
\end{equation}
where $q$ belongs to the upper half of the momentum plane. Because the
imaginary part of $q$ is positive, the exponential blow up
of $e^{ip_{\rm res}r/\hbar}$ mutates into the exponential damping of $e^{iqr/\hbar}$,
which allows us to obtain a well defined Gamow state through
Eq.~(\ref{inteGam08}). In Ref.~\cite{MONDRA}, this procedure was exemplified
through the resonances of the delta-shell potential located at $r=a$,
\begin{equation}
       V(\vec{x}) = \frac{\hbar ^2}{2ma} \lambda \delta (r-a) \, ,
        \label{exdspot}
\end{equation}
where $\lambda$ is a dimensionless constant that characterizes the 
strength of the potential. The $s$-wave resonant momenta of the delta-shell
potential satisfy~\cite{MONDRA}
\begin{equation}
          1+ \frac{\lambda \hbar}{p_{\rm res}a} e^{ip_{\rm res}a/\hbar} 
         \sin (p_{\rm res}a/\hbar) =0 \, .
        \label{condfores}
\end{equation}
The $s$-wave Gamow states of the delta-shell
potential are well known~\cite{MONDRA}:
\begin{equation}
       \langle \vec{x}|p_{\rm res}\rangle = \frac{u_0(r;p_{\rm res})}{r}
                       Y_0^0(\hat{x}) \, ,
      \label{gsdspos} 
\end{equation}
where
\begin{equation}
       u_0(r;p_{\rm res}) =   \left\{ \begin{array}{ll}
               N_{\rm res}\, e^{ip_{\rm res}a/\hbar} \sin (p_{\rm res} r/\hbar) \quad  
                               & 0<r<a \, , \\ [2ex]
              N_{\rm res} \, \sin (p_{\rm res} a/\hbar)  e^{ip_{\rm res}r/\hbar}        
     \quad &  a<r<\infty \, ,
        \end{array} 
      \right. 
	\label{rstate}
\end{equation}
and where
\begin{equation}
      N_{\rm res}= \left[   \frac{2\lambda}{a(1+\lambda - 2ip_{\rm res}a/\hbar)}
        \right]^{1/2}
      \label{normaconstant}
\end{equation}
is a normalization constant such that Eq.~(\ref{normalization}) 
holds. Hern\'andez and Mondrag\'on, through a lengthy calculation,
used Eq.~(\ref{inteGam08}) to obtain the momentum representation of the 
Gamow state~(\ref{gsdspos}).

In this section, we are going to obtain the same result, but using a 
much simpler method that works for potentials of finite range. When the 
potential is of finite range, the matrix element 
$\langle \pvec |V|p_{\rm res}\rangle$
is finite, even though the Gamow states blow up exponentially 
at infinity, and one can obtain the momentum representation of a Gamow
state through Eq.~(\ref{inteGamp0}) without having to perform the
analytical continuation of Eq.~(\ref{inteGam08}).

By using Eqs.~(\ref{inteGamp0}), (\ref{maelpte}), (\ref{andmbinpos}),
(\ref{exdspot}), (\ref{rstate}) and (\ref{condfores}) in quick succession, 
we can obtain the momentum
representation of the Gamow state for the $s$-wave resonances of
the delta-shell potential:
\begin{eqnarray}
         \langle \vec{p}|p_{\rm res}\rangle &=& 
                   \frac{1}{E_{\rm res}-E_p} \langle \vec{p}|V|p_{\rm res}\rangle
          \nonumber \\ 
         &=&
\frac{2m}{p_{\rm res}^2-p^2}   \sqrt{\frac{1}{mp}}
          \langle E|V|E_{\rm res}\rangle _0  \, Y_0^0(\hat{p})
           \nonumber \\
         &=& \frac{2m}{p_{\rm res}^2-p^2}
              \sqrt{\frac{1}{mp}}
              \int_0^{\infty} dr \, 
              \sqrt{\frac{2m}{\pi \hbar p }} \hat{j}_0(pr/\hbar)
               \frac{\hbar ^2}{2ma} \lambda \delta (r-a) u(r;p_{\rm res}) 
                  \, Y_0^0(\hat{p})
           \nonumber \\
         &=&  
\frac{1}{p_{\rm res}^2-p^2} 
              \sqrt{\frac{2}{\pi \hbar }} \frac{\hbar ^2}{a} \lambda
              \sin (pa/\hbar)
              u(a;p_{\rm res}) \, \frac{Y_0^0(\hat{p})}{p}
 \nonumber \\
         &=&  
\frac{1}{p_{\rm res}^2-p^2} 
              \sqrt{\frac{2}{\pi \hbar }} \frac{\hbar ^2}{a} \lambda
              \sin (pa/\hbar)
              N_{\rm res} \sin(p_{\rm res}a/\hbar) e^{ip_{\rm res}a/\hbar} 
               \, \frac{Y_0^0(\hat{p})}{p}
\nonumber \\
         &=&  
\frac{1}{p_{\rm res}^2-p^2} 
              \sqrt{\frac{2}{\pi \hbar }} \frac{\hbar ^2}{a} \lambda
              \sin (pa/\hbar)
              N_{\rm res} \frac{-p_{\rm res}a}{\hbar \lambda}
               \, \frac{Y_0^0(\hat{p})}{p}
     \nonumber \\
         &\equiv & \frac{\hat{u}_0(p;p_{\rm res})}{p}
               \, Y_0^0(\hat{p}) \, ,
          \label{gsmorepdelts}
\end{eqnarray}                
where
\begin{equation}
     \hat{u}_0(p;p_{\rm res}) = N_{\rm res} \sqrt{ \frac{2\hbar}{\pi}} 
                             \frac{p_{\rm res}}{p^2-p_{\rm res}^2}
                               \sin(pa/\hbar) \, .
          \label{momrepgsdo}
\end{equation}
is the ``radial'' part of the Gamow state in the momentum 
representation. Clearly, this function is square-integrable, and its norm
is equal to $\Gamma$,
\begin{equation}
      \int d^3p \, |\langle \vec{p}|p_{\rm res}\rangle |^2 =
      \int_0^{\infty} dp \, |\hat{u}_0(p;p_{\rm res})|^2 = \Gamma \, .
   \label{sinwfmrep}
\end{equation}

In the momentum representation, the normalization condition
of Eq.~(\ref{normalization}) is obtained by analytic continuation
from the upper-half of the momentum plane into the resonant pole~\cite{MONDRA},
\begin{eqnarray}
  \int_0^{\infty} \hat{u}_0(p;p_{\rm res})^2 \, dp  & \equiv &
  \left[ \int_0^{\infty} \hat{u}_0(p;q)^2 \, dp\right] _{q=p_{\rm res}}
  \nonumber \\
  &=& \left[ \int_0^{\infty}  N_{\rm res}^2 \frac{2\hbar}{\pi} 
                             \frac{q^2}{(p^2-q^2)^2}
                             \sin^2(pa/\hbar)
                             \, dp\right] _{q=p_{\rm res}}
  \nonumber \\
  &=&  N_{\rm res}^2 \frac{2\hbar}{\pi}\left[ \frac{-i\pi}{8q}
    (1+e^{2iaq/\hbar}(-1+2aq/\hbar))
             \right]_{q=p_{\rm res}}
                 \nonumber \\
  &=&  N_{\rm res}^2 \frac{2\hbar}{\pi} \frac{-i\pi}{8p_{\rm res}}
                 \frac{2ap_{\rm res}i}{\lambda \hbar}
                 (1+ \lambda  -2i a p_{\rm res}/\hbar)
            \nonumber \\     
   &=&  \frac{2\lambda}{a(1+\lambda - 2ip_{\rm res}a/\hbar)}
          \frac{2\hbar}{\pi} \frac{-i\pi}{8p_{\rm res}}
                 \frac{2ap_{\rm res}i}{\lambda \hbar}
                 (1+ \lambda  -2i a p_{\rm res}/\hbar)
          \nonumber \\     
   &=& 1 \,         
                 \label{normalmomentall}
\end{eqnarray}
where ${\rm Im}(q)>0$, and where we have used Eqs.~(\ref{momrepgsdo}),
(\ref{condfores}) and~(\ref{normaconstant}) in the second,
fourth and fifth steps, respectively. It is important to note that
the integral in the second step of Eq.~(\ref{normalmomentall}) needs to
be performed first when ${\rm Im}(q)>0$, and only after we can continue the
result to the resonant pole~\cite{MONDRA}.

Comparison of Eqs.~(\ref{sinwfmrep}) and~(\ref{normalmomentall}) shows
why, in general, $\Gamma$ is different from unity (and therefore why,
in general, $\overline{\Gamma}$ is different from $\gr$). The reason is 
that the normalization of Eq.~(\ref{normalmomentall}) utilizes
the square of the Gamow eigenfunction, whereas
Eq.~(\ref{sinwfmrep}) uses the absolute value squared.

\section{Conclusions}
\setcounter{equation}{0}
\label{sec:conclusions}

We have used a momentum basis to obtain the Gamow 
Golden Rule, and have shown that when the resonance is 
long-lived, the (exact) Gamow Golden Rule becomes the (approximate) Fermi 
Golden Rule. The Gamow 
Golden Rule provides the differential decay constant as the Lorentzian 
multiplied by the absolute value squared of the matrix element of the 
interaction, and can be used to describe both energy and angular
decay distributions. Phenomenologically, the Gamow Golden Rule is more suitable
to describe decay distributions than Fermi's Golden Rule, because 
the Gamow Golden Rule exhibits Lorentzian-like bumps, whereas the Fermi 
Golden Rule simply has a spike at the mean resonant energy $\er$. In addition,
the Gamow Golden Rule is more suitable to describe resonance interference
than Fermi's Golden Rule.

We have argued that the phase space factors of the Gamow Golden Rule should not 
include energy- and momentum-conserving delta functions, because 
experimentally the energy and the momentum of a resonance change on an 
event-by-event basis. We have shown that the momentum basis Golden Rule 
yields the same decay width as the angular momentum basis when the potential 
is spherically symmetric.

In spite of the fact that the Gamow states blow up exponentially 
at infinity in the position representation, we can obtain well-defined decay 
distributions because the Lippmann-Schwinger equation of the Gamow states 
yields well-defined, square-integrable Gamow states in the momentum 
representation. In general, one needs to use analytic continuation to
obtain the momentum representation of a Gamow state, but when the potential
is of finite range, we can obtain it directly
from the Lippmann-Schwinger equation of the Gamow states.

The Gamow Golden Rule is only relevant to describe the {\it irreversible} 
decay of a resonance~\cite{COHEN}. It does not describe other transitions
from an initial to a final state, such as Rabi oscillations. In addition,
although the momentum representation of a Gamow state can be 
interpreted as a decay distribution, the momentum representation of a 
regular wave function does not have the same interpretation. For example, 
one can always obtain the momentum representation of a bound state, but the 
ensuing probability distribution describes the outcomes of the measurement 
of momentum in the bound state, not the decay of a bound state into the 
continuum.

Finally, because the only approximation in the relativistic Golden 
Rule~(\ref{relafgr}) is that the resonance is long-lived, comparison of 
Eq.~(\ref{relafgr}) with the Gamow and the Fermi Golden Rules suggests that the 
exact relativistic Golden Rule can be obtained by replacing the delta
function with a relativistic Lorentzian. It is still an open question
whether that should be the expression for the exact
relativistic Golden Rule, though.

\section*{Acknowledgment}
\setcounter{equation}{0}
\label{sec:ack}

Stimulating conversations with Rodolfo Id Betan are gratefully 
acknowledged.


\begin{thebibliography}{99}
\bibitem{NOTE1} Often, Fermi's Golden Rule is written
(equivalently) in terms of the decay rate.
\bibitem{NOTE2} Usually, the expressions of the Gamow eigenfunctions
are simpler when they are expressed in terms of the energy
$E_{\rm res}$ or the wave number $k_{\rm res}=p_{\rm res}/\hbar$. However, to
stress that we are working with the momentum basis, we will
denote the Gamow state by $|p_{\rm res}\rangle$ instead of
$|E_{\rm res}\rangle$ or $|k_{\rm res}\rangle$.
\end{thebibliography}
\end{document}